\documentstyle[preprint,pra,aps,epsf]{revtex}

\begin{document}
\draft

\preprint{CfA 4463}

\title{Long-range interactions of lithium atoms}
\author{Zong-Chao Yan, A. Dalgarno, and J. F. Babb}
\address{
Harvard-Smithsonian Center for Astrophysics,\\
60 Garden Street, Cambridge, MA 02138
}

\maketitle
\begin{abstract}
The long-range interactions of two atoms, of an atom and a dielectric
wall, of an atom and a perfectly conducting wall, and of an atom
between two perfectly conducting walls are calculated, including the
effects of retardation, for Li using dynamic polarizabilities
determined from highly correlated, variationally determined wave
functions.
\end{abstract}
\narrowtext
\section{INTRODUCTION}
Long-range interactions between two ground state atoms and between a
ground state atom and a surface are now measured using lasers and cold
atoms or atomic beams.  Photoassociation spectroscopy has yielded
strict limits on the values of coefficients of dispersion forces
between two Rb atoms~\cite{GarCliMil95},
two Na atoms~\cite{JonJulLet96}, and two
Li atoms~\cite{McAAbrHul96} 
in their ground states.
Spectroscopy combined with deflection of alkali-metal atomic beams
near surfaces~\cite{SukBosCho93} or reflection of Na atoms from
surfaces in atomic fountains~\cite{KasMolRii91,LanCouLab96} have made
it possible to measure the coefficients of atom-surface forces.  The
experiments are consistent with theoretical models for the interaction
potentials, but accurate theoretical estimations of the potentials
remain elusive for atoms other than H and He.  The effects of
retardation, due to the finite speed of light, cause the potentials to
become weaker, approaching simple power laws for asymptotically large
distances.  The advent of highly-correlated basis sets for Li using
multiple nonlinear variational parameters~\cite{YanDra95} makes it
possible to perform well-converged calculations of the dynamic
electric polarizability functions, thereby enabling, as we will show,
precise evaluation of long-range interaction potentials, including
retardation, for two Li atoms, for a Li atom and a dielectric or
perfectly conducting wall, and for a Li atom between two perfectly
conducting walls.

\section{FORMULATION}
\subsection{Atom-atom interactions}

The effect of retardation on the long-range induced dipole
interactions of two atoms was first investigated by Casimir and
Polder~\cite{CasPol48} and the effects on higher induced multipoles by
Au and Feinberg~\cite{AuFei72}, Jenkins, Salam, and
Thirunamachandran~\cite{JenSalThi94}, and Power and
Thirunamachandran~\cite{PowThi96}. In this paper, the dipolar-dipolar
and dipolar-quadrupolar interactions are considered, the higher
multipolar interactions being negligible. We use the expression for
the retarded dipolar-quadrupolar interaction of Power and
Thirunamachandran~\cite{PowThi96} which differs from the approximate
expression obtained by Au and Feinberg~\cite{AuFei72}.

The interaction potential between two like atoms including the effects
of retardation can be written~\cite{CasPol48,PowThi96}
\begin{equation}
V (R) = - \frac{C_6f_6 (R)}{R^6}- \frac{C_8f_8 (R)}{R^8},
\end{equation}
where
\begin{equation}
\label{C6}
C_6 = \frac{3}{\pi}G(1,1)\,,
\end{equation}
\begin{equation}
\label{C8}
C_8 = 
\frac{15}{\pi}G(1,2)\,,
\end{equation}
with 
\begin{equation}
G(l_{\rm a},l_{\rm b}) = \int_{0}^{\infty}
\alpha_{l_{\rm a}}(i\omega)
\alpha_{l_{\rm b}}(i\omega)d\omega\,
\end{equation}
and the retardation coefficients are
\begin{equation}
\label{f6}
f_6 (R)  = 
 \frac{1}{\pi C_6} \int_0^\infty 
       d\omega \exp (-2\alpha_{\rm fs}\omega R)
       \alpha_1^2 (i\omega) P_{11}(\omega\alpha_{\rm fs} R),
\end{equation}
where
\begin{equation}
P_{11} (x) = 
 x^4 + 2x^3 + 5x^2 + 6x + 3 
\end{equation}
and 
\begin{equation}
\label{f8}
f_8 (R) = 
 \frac{1}{3\pi C_8} \int_0^\infty 
       d\omega \exp (-2\alpha_{\rm fs}\omega R)
       \alpha_1 (i\omega)\alpha_2 (i\omega) P_{12}(\omega\alpha_{\rm fs} R),
\end{equation}
where
\begin{equation}
P_{12} (x) = 
 \case{1}{2}x^6 + 3x^5 + \case{27}{2}x^4 + 42x^3 +81x^2  +90x +45 ,
\end{equation}
and $\alpha_{\rm fs}=1/137.035\,989\,5$ is the fine structure constant.  We use
atomic units throughout.

The functions $\alpha_{l}(i\omega)$ appearing in
(\ref{C6})--(\ref{f8}) are the dynamic electric multipole
polarizability functions at imaginary frequency defined by expressions
(6)--(9) of~\cite{YanBabDal96}.

The retardation coefficients are dimensionless and can be expanded for
small $R$ as
\begin{equation}
C_6 f_6 (R) \sim C_6 - \alpha_{\rm fs}^2R^2 W_4 
\end{equation}
with 
\begin{equation}
\label{W4}
W_4 =  \frac{1}{\pi} \int_0^\infty 
       d\omega\;\omega^2 \alpha_1^2 (i\omega)
\end{equation}
and 
\begin{equation}
\label{f8-small}
C_8 f_8 (R) \sim C_8 - \alpha_{\rm fs}^2R^2 W_6 ,
\end{equation}
with 
\begin{equation}
\label{W6}
W_6 = \frac{3}{\pi} \int_0^\infty 
       d\omega\;\omega^2 \alpha_1 (i\omega)\alpha_2 (i\omega) .
\end{equation}

The coefficients $W_4$ and $W_6$ can also be derived from an analysis
in perturbation theory of the orbit-orbit term arising from the Breit
interaction in the Pauli approximation~\cite{MeaHir66}.  Expanding
Power and Thirunamachandran's result for $-C_8f_8 (R)$ for small $R$,
according to (\ref{f8-small}), we find a value of $W_6$ a factor of
$\case{3}{2}$ times larger than the value of $W_6$ resulting from the
theory of Au and Feinberg (see Eq.~(4.21) of~\cite{MarBabDal94}). This
resolves the discrepancy of $\case{3}{2}$ found in~\cite{MarBabDal94},
between the value of $W_{LL,4;2}$ for H from Johnson, Epstein, and
Meath~\cite{JohEpsMea67}, who evaluated terms from the Breit
interaction in the Pauli approximation~\cite{MeaHir66}, and the value
of $W_6$ evaluated for H in~\cite{MarBabDal94} using the Au and
Feinberg formulation.

For asymptotically large $R$, the retardation coefficients have the
limits
\begin{equation}
f_6 (R) \rightarrow \frac{23}{4\pi\alpha_{\rm fs} R}
\frac{\alpha_1^2 (0)}{C_6},
 \qquad R\rightarrow \infty ,
\end{equation}
and 
\begin{equation}
f_8 (R) \rightarrow \frac{531}{16\pi\alpha_{\rm fs} R}
\frac{\alpha_1 (0)\alpha_2 (0)}{C_8},
 \qquad R\rightarrow \infty .
\end{equation}

\subsection{Atom-wall interactions}
Expressions for the interaction potential of an atom and a dielectric
wall, including the effects of retardation, have been given by
Dzyaloshinskii, Lifshitz, and Pitaevskii~\cite{DzyLifPit61},
Parsegian~\cite{Par74}, and Tikochinsky and Spruch~\cite{TikSpr93a}.
For a wall with a dielectric constant $\epsilon$, the potential can be
written~\cite{TikSpr93a}
\begin{equation}
\label{AtD}
V_{{\rm At}D} (R,\epsilon) =
  -\frac{\alpha_{\rm fs}^3}{2\pi} \int_0^\infty d\xi \xi^3 \alpha_1 (i\xi)
     \int_1^\infty dp \exp (-2\xi R p\alpha_{\rm fs}) H (p,\epsilon) ,
\end{equation}
where
\begin{equation}
H (p,\epsilon) = \frac{s-p}{s+p} 
        + (1-2p^2)\frac{s-\epsilon p}{s+\epsilon p} 
\end{equation}
and 
\begin{equation}
s = (\epsilon - 1 +p^2)^{1/2}.
\end{equation}
We follow the notation of~\cite{SprTik93} and the subscripts At,
$D$, and $M$ denote, respectively, an atom, a dielectric wall, and a
perfectly conducting ({\em i.e.\/} metal) wall.

For asymptotically large distances, 
\begin{equation}
\label{dzyal}
V_{{\rm At}D} (R,\epsilon) \rightarrow
    V^\infty_{{\rm At}D} (R,\epsilon) = -\frac{K_4}{R^4} 
         \frac{\epsilon -1}{\epsilon +1} \phi (\epsilon),
\end{equation}
where 
\begin{equation}
\label{casimir}
K_4 =  3 \alpha_1 (0)/  (8\pi\alpha_{\rm fs}) = 16.36\, \alpha_1 (0)  
\end{equation}
and 
\begin{equation}
\label{phi}
\phi(\epsilon) = \frac{\epsilon+1}{2 (\epsilon-1)} 
              \int_0^\infty \frac{dp}{(p+1)^4} H (p+1,\epsilon).
\end{equation}
Direct integration of (\ref{phi}) yields
\begin{equation}
\phi (\epsilon) = 
  \frac{\epsilon +1}{\epsilon -1}
   \left[ \frac{1}{3} + \epsilon + 
   \frac{4- (\epsilon+1)\epsilon^{1/2}}
         {2 (\epsilon-1)} + A(\epsilon)+B(\epsilon) \right] ,
\end{equation}
where
\begin{equation}
\label{A}
A (\epsilon) = -\frac{{\rm Arcsinh}[(\epsilon-1)^{1/2}]}
          {2 (\epsilon-1)^{3/2}}
               [1 + \epsilon+2\epsilon (\epsilon-1)^2]
\end{equation}
and 
\begin{equation}
B (\epsilon) = 
            \frac{\epsilon^2}{(\epsilon+1)^{1/2}}
        [{\rm Arcsinh}(\epsilon^{1/2})-{\rm Arcsinh}(\epsilon^{-1/2})] ,
\end{equation}
in agreement with 
Dzyaloshinskii {\em et al.\/}~\cite{DzyLifPit61}.
Approximations to $V^\infty_{{\rm At}D} (R,\epsilon)$ will be
considered in Sec.~\ref{sec:calc} below.

The potential for the interaction of an atom and a perfectly
conducting wall follows by letting $\epsilon \rightarrow \infty$ in
(\ref{AtD}) giving~\cite{CasPol48}
\begin{equation}
\label{AtM}
V_{{\rm At}M} (R) \equiv  
V_{{\rm At}D} (R,\infty) 
     = -\frac{C_3f_3 (R)}{R^3} ,
\end{equation}
where the coefficient is 
\begin{equation}
\label{C3}
C_3 = \frac{1}{4\pi} \int_0^\infty \,d\omega \alpha_1 (i\omega) ,
\end{equation}
and the retardation coefficient is
\begin{equation}
\label{f3}
f_3 (R) = \frac{1}{8 C_3 \pi \alpha_{\rm fs} R} 
       \int_0^\infty dx\; e^{-x}\alpha_1 (ix/2\alpha_{\rm fs} R) 
         [\case{1}{2}x^2+x+1]  .
\end{equation}
Eq.~(\ref{f3}) approaches for asymptotically large distances the form
\begin{equation}
f_3 (R) \rightarrow \frac{3}{8\pi}\frac{\alpha_1 (0)}{\alpha_{\rm fs}C_3} ,
\end{equation}
giving 
\begin{equation}
\label{asymp}
V_{{\rm At}M}(R) 
    \rightarrow V^\infty_{{\rm At}M} (R) \equiv -K_4/R^4.
\end{equation}

The interaction potential for an atom between two
parallel, perfectly conducting walls has been given
by Barton~\cite{Bar87a} and by Zhou and Spruch~\cite{ZhoSpr95}.
It can be expressed as 
\begin{equation}
\label{MAtM}
V_{M{\rm At}M} (z,L) =  T_2 (L) -  T_1 (z,L) ,
\end{equation}
where 
\begin{equation}
T_1 (z,L) = \frac{1}{\pi L^3} \int_0^\infty
     dt \frac{t^2 \cosh (2zt/L)}{\sinh t} 
     \int_0^{t/\alpha_{\rm fs} L}  ds \alpha_1 (is)  
\end{equation}
and
\begin{equation}
T_2 (L) = \frac{\alpha_{\rm fs}^2}{\pi L} 
   \int_0^\infty  ds \, s^2\alpha_1 (is) 
    \int_{\alpha_{\rm fs} Ls}^\infty  dt \frac{e^{-t}}{\sinh t}  ,
\end{equation}
where $L$ is the interwall distance and $z$
is the distance of the atom from the midpoint.
For small values of $L$, the potential is~\cite{ZhoSpr95}
\begin{equation}
V_{M{\rm At}M} (z,L) \rightarrow -\frac{4}{L^3} T (z/L) C_3,
\end{equation}
where 
\begin{equation}
T (z/L) = \int_0^\infty dt 
    \frac{t^2\cosh (2tz/L)}{\sinh t} 
\end{equation}
and $C_3$ is defined in (\ref{C3}).  For asymptotically large values
of $L$, the potential is
\begin{equation}
V_{M{\rm At}M}^\infty (z,L) = 
   \frac{\pi^3\alpha_1 (0)}{\alpha_{\rm fs} L^4}
  \left [ \frac{1}{360} 
       -\frac{3-2\cos^2 (\pi z/L)}{8\cos^4 (\pi z/L)} \right].
\end{equation}

\section{CALCULATIONS}
\label{sec:calc}

The calculations of the wave functions and the polarizability response 
functions have been described previously in, respectively,
~\cite{YanDra95} and~\cite{YanBabDal96}.  We briefly summarize the
procedures.

The basis set for the lithium atom is constructed in Hylleraas
coordinates \cite{YanDra95}
\begin{eqnarray}
\{\phi_{t,\mu_t} (\alpha_t,\beta_t,\gamma_t) &=& r_1^{j_1}\,r_2^{j_2}
\,r_3^{j_3}\,r_{12}^{j_{12}}\,r_{23}^{j_{23}} \,r_{31}^{j_{31}}
\mbox{}e^{-\alpha_t r_1-\beta_t r_2-\gamma_t r_3}\}\,,
\label {eq:a23}
\end{eqnarray}
where $\mu_t$ denotes a sextuple of integer powers $j_1$, $j_2$,
$j_3$, $j_{12}$, $j_{23}$, and $j_{31}$, index $t$ labels different
sets of nonlinear parameters $\alpha_t$, $\beta_t$ and $\gamma_t$.
Except for some truncations, all terms are included such that
\begin{eqnarray}
j_1+j_2+j_3+j_{12}+j_{23}+j_{31} &\leq& \Omega\,.
\label {eq:a24}
\end{eqnarray}
The wave function is expanded from the multiple basis sets
\begin{eqnarray}
\Psi ({\bf r}_1,{\bf r}_2,{\bf r}_3) &=& {\cal A} \sum_{t}
\sum_{\mu_t}a_{t,\mu_t}\phi_{t,\mu_t} (\alpha_t,\beta_t,\gamma_t)
\nonumber\\
&&\mbox{}\times ({\rm angular\ function})({\rm spin\ function})\,.
\label {eq:aa25}
\end{eqnarray}
A complete optimization is performed with respect to all the nonlinear
parameters. The screened hydrogenic wave function is also included
explicitly in the basis set.

The dynamic polarizabilities are evaluated using effective oscillator
strengths and transition energies obtained from the diagonalization of
the Hamiltonian in a basis set of $S$ symmetry for the ground state
and of $P$ and $D$ symmetry, respectively, for the intermediate states
corresponding to the dipole and quadrupole polarizabilities.  The
basis sets were the size 919 set from~\cite{YanBabDal96} for the $S$
symmetry and the size 1846 sets from~\cite{YanBabDal96} for the $P$
and the $D$ symmetries.  A detailed discussion of the evaluation of
$\alpha_l(i\omega)$ can be found in \cite{YanBabDal96}.  The static
polarizabilities have the values $\alpha_1 (0)=164.111 (2)$ and
$\alpha_2 (0)=1\,423.266 (5)$~\cite{YanBabDal96}.

\begin{figure}[p]
\epsfxsize=1.\textwidth \epsfbox{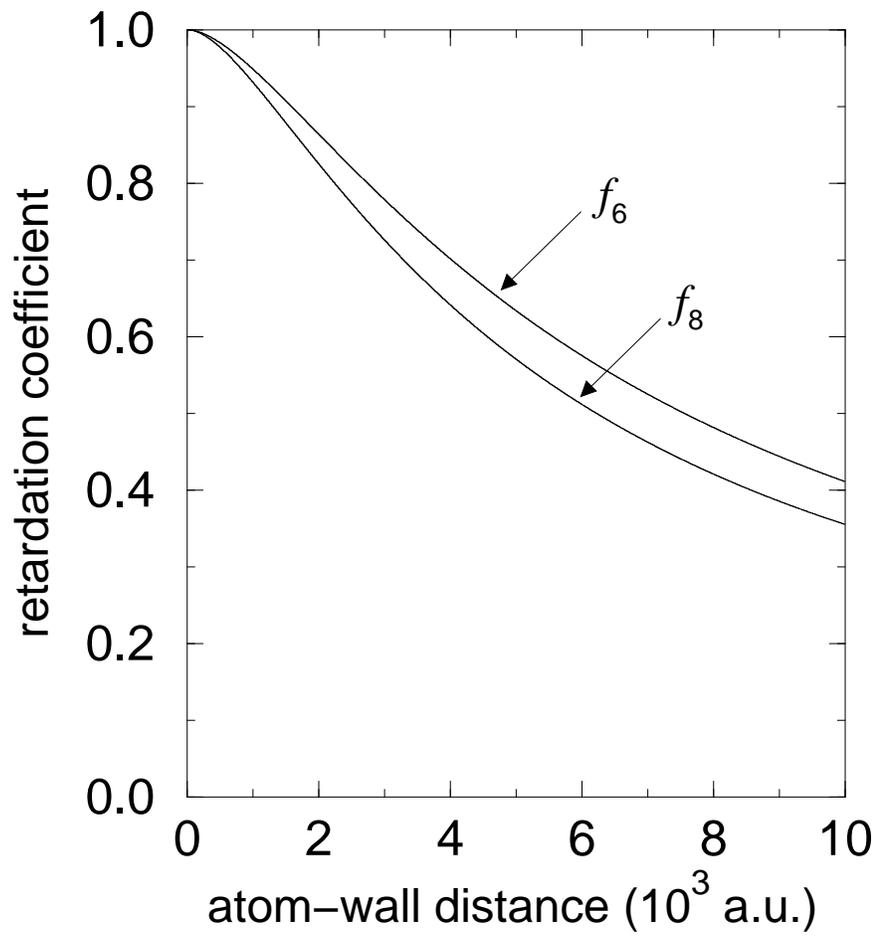}
\caption{Dimensionless retardation coefficients
$f_6 (R)$ and $f_8 (R)$  for two Li atoms.
\label{f6-fig}}
\end{figure}

Values of the coefficient $W_4$ for two Li atoms have been determined
by Margoliash and Meath~\cite{MarMea78} and by Easa and
Shukla~\cite{EasShu83}.  Using our functions $\alpha_1 (i\omega)$ and
$\alpha_2 (i\omega)$, we determined the coefficients $W_4$ and $W_6$
using, respectively, (\ref{W4}) and (\ref{W6}), and the results are
compared with previous results in Table~\ref{W-table}.  We also
calculated the coefficients $f_6 (R)$ and $f_8 (R)$ using (\ref{f6})
and (\ref{f8}) at various values of $R$.  The results are given in
Table~\ref{f-table} and Fig.~\ref{f6-fig}. The values of the
dipole-dipole potential $-C_6f_6 (R)/R^6$
are in agreement with, but are more accurate than, those given in
Ref.~\cite{MarBabDal94}, calculated using a model potential method.
The values of the dipole-quadrupole potential $-C_8f_8 (R)/R^8$ 
replace those given in
Ref.~\cite{MarBabDal94}, which were calculated using the expression of
Au and Feinberg as discussed above.
The dipole-quadrupole potential is usually of secondary importance
due to its $1/R^8$ power law behavior.

\begin{figure}[p]
\epsfxsize=1.\textwidth \epsfbox{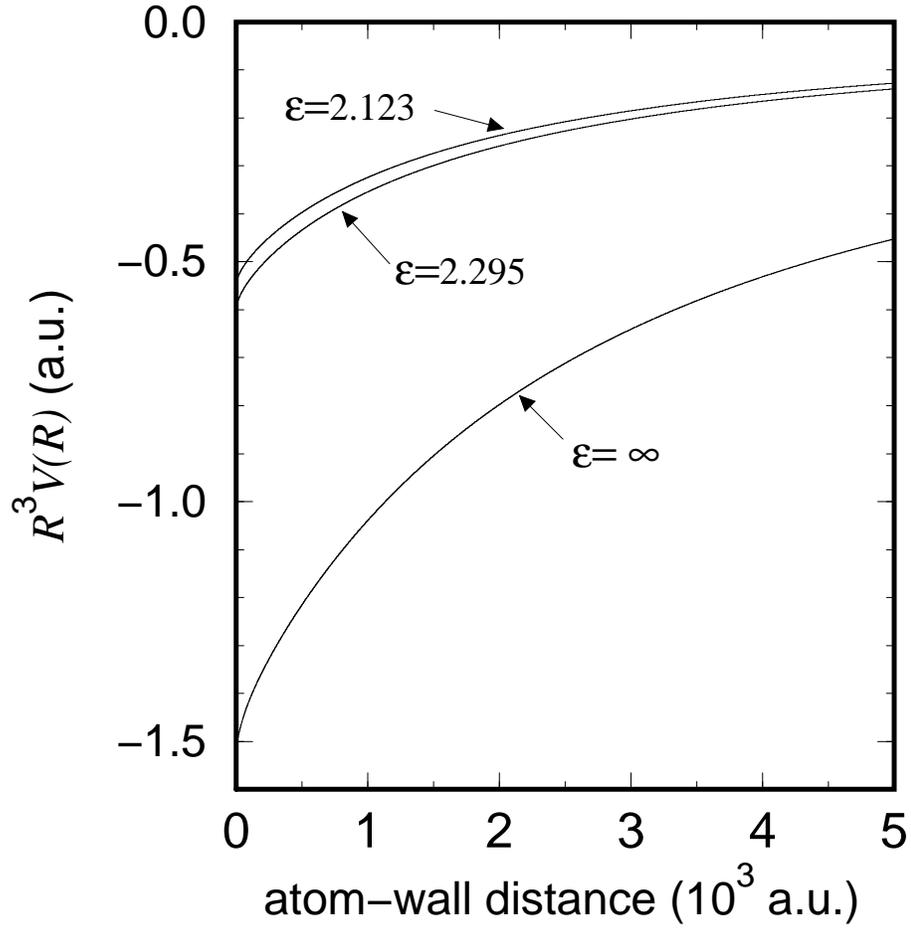}
\caption{
Values of $R^3V_{{\rm At}D}(R,\epsilon)$,
for
values of $\epsilon$ corresponding
to fused silica $(\epsilon=2.123)$, BK-7 glass  $(\epsilon=2.295)$,
and 
for a perfectly conducting wall  $(\epsilon=\infty)$.
\label{wall-fig}}
\end{figure}

Using the polarizability function $\alpha_1 (i\omega)$ we evaluated
$V_{{\rm At}D} (R,\epsilon)$ for values of $\epsilon= 2.123$ and
$2.295$ corresponding to, respectively, fused silica and BK-7
glass. The values are listed in Table~\ref{glass-table} and
illustrated in Fig.~\ref{wall-fig} for values of $R$ up to $5000a_0$.
For larger values of $R$, the potential can be obtained from
(\ref{dzyal}). The values of $\phi (\epsilon)$ from our calculations
are listed in Table~\ref{phi-table} and they are in agreement with the
representative values given in Fig.~10 of Ref.~\cite{DzyLifPit61}.

\begin{figure}[p]
\epsfxsize=1.\textwidth \epsfbox{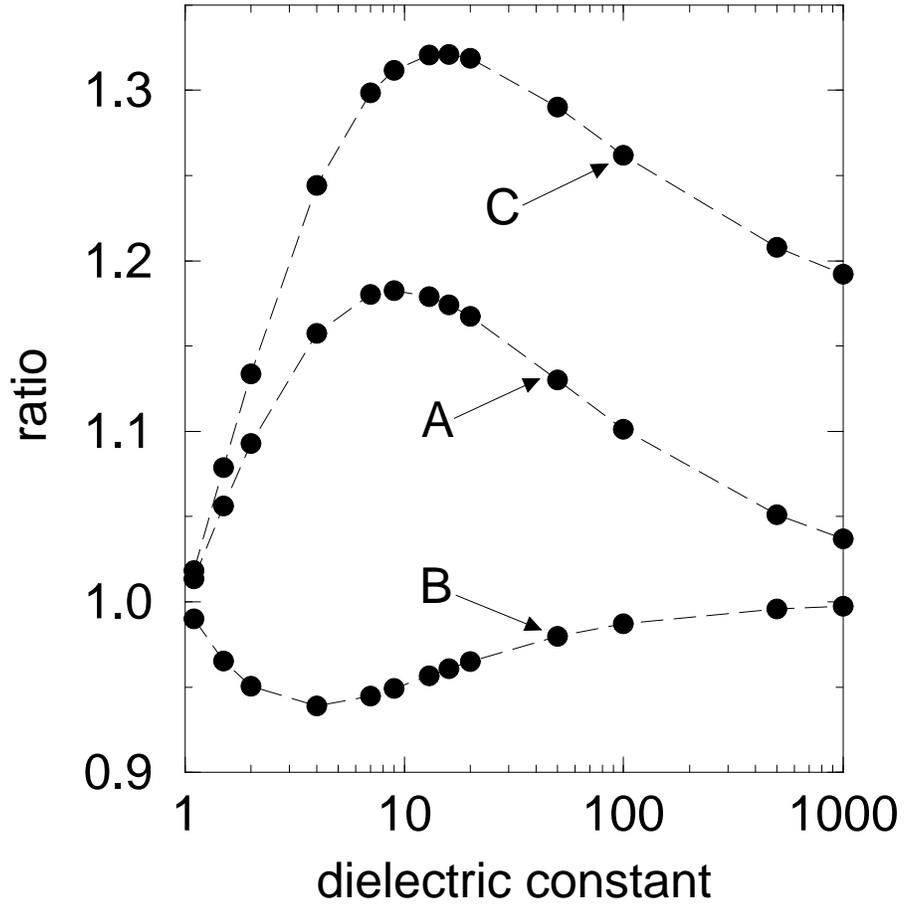}
\caption{Ratio of several approximations given
by Spruch  and Tikochinsky to the 
the exact atom-wall potential calculated in the present work.
The symbols A, B, and C represent, respectively,
values from
Eqs.~(\ref{prime}), (\ref{double-prime}), and (\ref{triple-prime}).
\label{prime-fig}}
\end{figure}

Three approximations $V'_{{\rm At}D}$, $V''_{{\rm At}D}$, and
$V'''_{{\rm At}D}$ for $V^\infty_{{\rm At}D}$ were obtained by
Spruch and Tikochinsky by imposing the
requirements that the interaction be exact for
$\epsilon \approx \infty$ and for $\epsilon \approx 1$, 
see Eqs.~(4.5), (4.9), and (4.12)
of~\cite{SprTik93}.  Expressing the approximations as ratios to the
exact potential at very large distances, we have
\begin{equation}
\label{prime}
\frac{V'_{{\rm At}D}}{V^\infty_{{\rm At}D} }
          = \frac{F(\epsilon)}{\epsilon + \case{37}{23}}  ,
\end{equation}
\begin{equation}
\label{double-prime}
\frac{V''_{{\rm At}D}}{V^\infty_{{\rm At}D} }
          = \frac{F(\epsilon)}{\epsilon + \case{30}{23}\epsilon^{1/2}
                      +\case{7}{23}}  ,
\end{equation}
and 
\begin{equation}
\label{triple-prime}
\frac{V'''_{{\rm At}D}}{V^\infty_{{\rm At}D} }
          = \frac{\case{23}{20}F(\epsilon)}{\epsilon+2} ,
\end{equation}
where 
\begin{equation}
F (\epsilon ) \equiv \frac{\epsilon+1}{\phi (\epsilon)} .
\end{equation}
We calculated the ratios appearing in
(\ref{prime})--(\ref{triple-prime}) using our values of $\phi
(\epsilon)$ and the results are presented in Fig.~\ref{prime-fig}.
Our results indicate that the second approximation defined by
(\ref{double-prime}) is the most accurate, differing by about 6\% at
most from the exact value of the potential.  The third approximation
was developed for small values of $\epsilon$ where it is seen to be
somewhat less accurate than the second approximation.

The interaction potential for a Li atom and a perfectly conducting
wall was evaluated from (\ref{AtM})--(\ref{f3}).  The value for the
coefficient $C_3$ is in excellent agreement with previous determinations,
listed in Table~\ref{C3-table},  particularly with
those calculated from the alternative expression
\begin{equation}
\label{C3-direct}
C_3 = \frac{1}{12} 
   \left \langle 0 \left| \left(\sum_{i=1}^N{\bf r}_i \right)^2
          \right| 0 \right\rangle ,
\end{equation}
which follows from integration of (\ref{C3}), where $N$ is the number
of electrons.  Note that only the ground state wave function is
required to evaluate~(\ref{C3-direct}). For Table~\ref{C3-table} we
used expectation values given by King~\cite{Kin89} and Yan and
Drake~\cite{YanDra95}.  The values of $R^3 V(R)$ were calculated from
(\ref{f3}) and values are listed in Table~\ref{glass-table}.  The
present calculations of the potential values are in agreement with,
but are more accurate than, those given in Ref.~\cite{MarDalBab96}.

\begin{figure}[p]
\epsfxsize=1.\textwidth \epsfbox{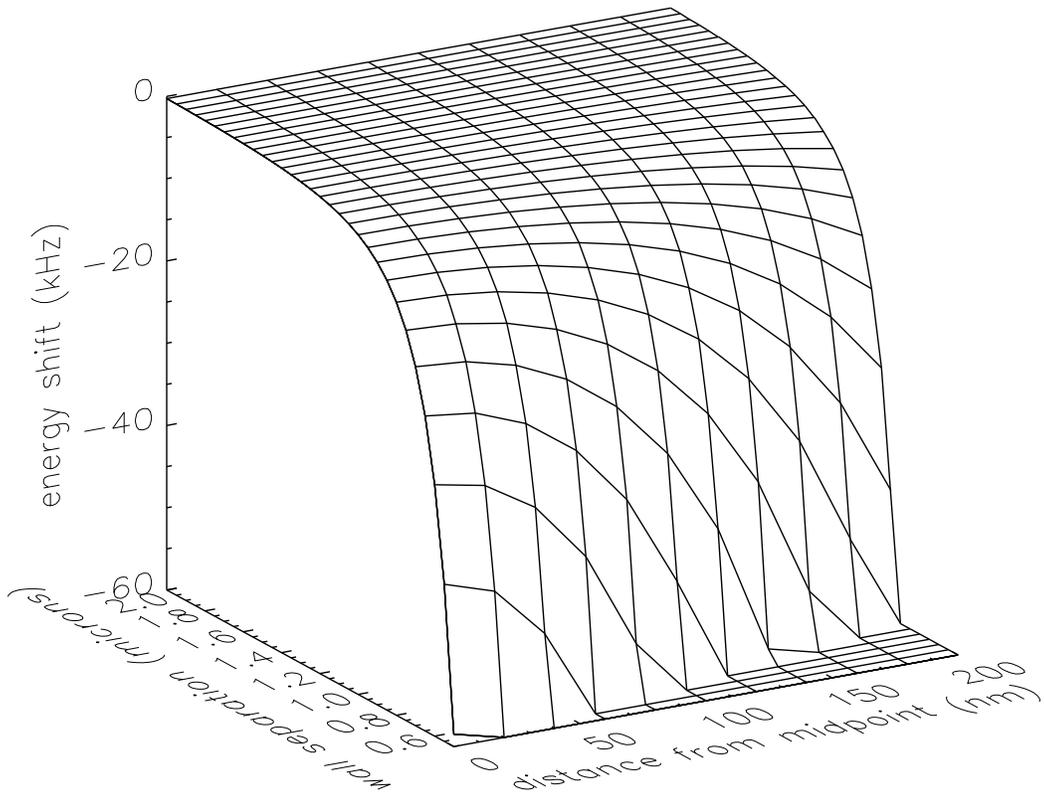}
\caption{The energy shift arising from the wall-atom-wall
potential for various values of the wall separation $L$ and the
distance of the atom from the midpoint $z$.  Only the values for $z>0$
are shown as the energy shift is symmetric about the $z=0$ plane.
\label{waw-fig}}
\end{figure}

The potential $V_{M{\rm At}M} (z,L)$ was evaluated using (\ref{MAtM})
for a range of wall separations $L$ and distances $z$ of the atom from
the midpoint. Values of the energy shift arising from the potential
for values of $L$ and $z$ that might be realized in an experiment are
given in Fig.~\ref{waw-fig}.

The expressions involving dielectric walls in this paper were obtained
under the approximation $\epsilon (\omega) \approx \epsilon (0)\equiv
\epsilon$, where $\epsilon (\omega)$ is the frequency-dependent
dielectric function of the wall.  This is an excellent approximation
for $R\sim\infty$, but at smaller $R$ it could lead to significant
error if resonances play a role.

\acknowledgements
We thank the referee for suggesting some improvements to the 
original version of the manuscript.
The Institute for Theoretical Atomic
and Molecular Physics is supported by a grant from the National
Science Foundation to the Smithsonian Institution and Harvard
University. Z.-C. Y. is also supported by the Natural Sciences
and Engineering Research Council of Canada and A.~D. by
the Office of Basic Energy Science, U.~S. Department of Energy.

\begin{table}
\begin{center}
\caption{
The coefficients $W_4$ and $W_6$ for two Li atoms.
Numbers in
parentheses represent theoretical uncertainty due to the finite basis
set size.}
\label{W-table}
\begin{tabular}{lll}
   $W_4$ & $W_6$  & Reference\\
\hline
 3.214(2)    &  219.9(2)       & Present\\
 2.9312      &               & Easa and Shukla~\protect\cite{EasShu83}\\
 3.233       &               & Margoliash and Meath~\protect\cite{MarMea78}\\
\end{tabular}	       				   
\end{center}
\end{table}
\clearpage
\begin{table}
\begin{center}
\caption{
The coefficient $C_3$ for the Li atom-wall interaction calculated in
the present work compared to values calculated from (\ref{C3-direct}),
(direct), using matrix elements given by various authors, or from
pseudo oscillator strength distribution data of dimension $M$
tabulated by various authors (osc. str.).  }
\label{C3-table}
\begin{tabular}{lll}
   $C_3$  & Method & Reference\\
\hline
   $1.518 (2)$            & osc. str.   & Present\\
   $1.518\,000\,51 (3)$   & direct  & Yan and Drake~\protect\cite{YanDra95}\\
   $1.518\,000$           & direct  &
                             King~\protect\cite{Kin89}\\
   $1.49$        & osc. str. $(M=11)$ & 
                      Stacey and Dalgarno~\protect\cite{StaDal68}\\
   $1.52$        & osc. str. $(M=10)$ &
                      Margoliash and Meath~\protect\cite{MarMea78}\\
\end{tabular}	       				   
\end{center}
\end{table}
\clearpage
\begin{table}
\begin{center}
\caption{
The dimensionless retardation coefficients $f_6 (R)$ and $f_8 (R)$ for
the atom-atom interaction.  The dispersion coefficients $C_6$ and
$C_8$ from~\protect\cite{YanBabDal96} are also given.}
\label{f-table}
\begin{tabular}{lcc}
     &  \multicolumn{1}{c}{$C_6$} &\multicolumn{1}{c}{$C_8$} \\
              &  $1\,393.39 (16)$ & $83\,425.8 (4.2)$  \\
    $R$ & $f_6 (R)$ & $f_8 (R)$  \\
\hline
     15 & 1.0000 & 1.0000   \\  
     20 & 1.0000 & 0.9999   \\  
     25 & 0.9999 & 0.9999   \\  
     30 & 0.9999 & 0.9999   \\  
     50 & 0.9997 & 0.9997   \\  
     70 & 0.9995 & 0.9994   \\  
    100 & 0.9991 & 0.9988   \\  
    150 & 0.9980 & 0.9974   \\  
    200 & 0.9966 & 0.9955   \\  
    250 & 0.9950 & 0.9933   \\  
    300 & 0.9931 & 0.9907   \\  
    500 & 0.9833 & 0.9775   \\  
    700 & 0.9708 & 0.9608   \\  
   1000 & 0.9489 & 0.9319   \\  
   1500 & 0.9076 & 0.8791   \\  
   2000 & 0.8641 & 0.8256   \\  
   2500 & 0.8208 & 0.7743   \\  
   3000 & 0.7789 & 0.7263   \\  
   5000 & 0.6341 & 0.5709   \\  
   7000 & 0.5253 & 0.4627   \\  
  10000 & 0.4113 & 0.3555   \\  
  15000 & 0.2970 & 0.2528   \\  
  20000 & 0.2304 & 0.1947   \\  
  25000 & 0.1875 & 0.1579   \\  
  30000 & 0.1578 & 0.1326   \\  
  50000 & 0.0961 & 0.0805   \\  
  70000 & 0.0689 & 0.0577   \\  
 100000 & 0.0484 & 0.0405   \\  
\end{tabular}	       				   
\end{center}
\end{table}
\begin{table}
\begin{center}
\caption{
Values of $-R^3V_{{\rm At}D}(R,\epsilon)$,
where $V_{{\rm At}D}(R,\epsilon)$ is
the atom-wall potential, for
values of $\epsilon$ corresponding
to fused silica and BK-7 glass in, respectively,
cols.~2 and 3,
and in col.~4 values of $-R^3V_{{\rm At}M}(R)$
for a perfectly conducting wall.}
\label{glass-table}
\begin{tabular}{lccc}
 \multicolumn{1}{c}{ } & \multicolumn{1}{c}{Fused silica}
           & \multicolumn{1}{c}{BK-7 glass} 
           & \multicolumn{1}{c}{perfect}\\
    $R$ &  $\epsilon=2.123$ & $\epsilon=2.295$ & $\epsilon=\infty$\\
\hline
     10 & 0.5360 & 0.5859  & 1.5007 \\ 
     15 & 0.5323 & 0.5819  & 1.4937 \\ 
     20 & 0.5289 & 0.5782  & 1.4871 \\ 
     25 & 0.5259 & 0.5749  & 1.4810 \\ 
     30 & 0.5230 & 0.5717  & 1.4753 \\ 
     50 & 0.5130 & 0.5608  & 1.4551 \\ 
     70 & 0.5045 & 0.5515  & 1.4380 \\ 
    100 & 0.4933 & 0.5392  & 1.4157 \\ 
    150 & 0.4772 & 0.5215  & 1.3836 \\ 
    200 & 0.4629 & 0.5060  & 1.3551 \\ 
    250 & 0.4500 & 0.4919  & 1.3289 \\ 
    300 & 0.4381 & 0.4788  & 1.3042 \\ 
    500 & 0.3974 & 0.4344  & 1.2160 \\ 
    700 & 0.3644 & 0.3983  & 1.1393 \\ 
   1000 & 0.3244 & 0.3546  & 1.0398 \\ 
   1500 & 0.2741 & 0.2996  & 0.9049 \\ 
   2000 & 0.2368 & 0.2589  & 0.7981 \\ 
   2500 & 0.2081 & 0.2276  & 0.7118 \\ 
   3000 & 0.1853 & 0.2026  & 0.6409 \\ 
   5000 & 0.1276 & 0.1395  & 0.4526 \\ 
\end{tabular}	       				   
\end{center}
\end{table}
\begin{table}
\begin{center}
\caption{
The dimensionless function $\phi (\epsilon)$.}
\label{phi-table}
\begin{tabular}{ll}
   $\epsilon$ & $\phi(\epsilon)$  \\
\hline
      1 &    $\case{23}{30}$ \\
  1.5   & 0.761364  \\ 
      2 & 0.760757  \\ 
  2.123 & 0.760970  \\
  2.295 & 0.761425  \\
      4 & 0.770171  \\ 
      7 & 0.787334  \\ 
      9 & 0.797062  \\ 
     13 & 0.812791  \\ 
     16 & 0.822186  \\ 
     20 & 0.832501  \\ 
     50 & 0.874337  \\ 
    100 & 0.902534  \\ 
    500 & 0.950261  \\ 
   1000 & 0.963647  \\ 
   5000 & 0.982986  \\ 
  10000 & 0.987836  \\ 
  50000 & 0.994478  \\ 
  $1\times 10^{11}$ & 0.999996  \\ 
\end{tabular}	       				   
\end{center}
\end{table}

\clearpage


\end{document}